# Fifty Years of $^3$He-rich Events

*Donald V. Reames*

Institute for Physical Science and Technology, University of Maryland, College Park, MD, USA

dvreames@gmail.com

**Abstract** The early 1970s saw a new and surprising feature in the composition of solar energetic particles (SEPs), resonant enhancements up to 10,000-fold in the ratio $^3$He/$^4$He that could even make $^3$He dominant over H in rare events. It was soon learned that these events also had enhancements in the abundances of heavier elements, such as a factor of ~10 enhancements in Fe/O, which was later seen to be part of a smooth increase in enhancements vs. mass-to-charge ratio $A/Q$ from H to Pb, rising by a factor of ~1000. These events were also associated with streaming 10 – 100 keV electrons that produce type III radio bursts. In recent years we have found these "impulsive" SEP events to be accelerated in islands of magnetic reconnection from plasma temperatures of 2 – 3 MK on open field lines in solar jets. Similar reconnection on closed loops traps the energy of the particles to produce hot (>10 MK), bright flares. Sometimes impulsive SEP intensities are boosted by shock waves when the jets launch fast coronal mass ejections. No single theory yet explains both the sharp resonance in $^3$He and the smooth increase up to heavier elements; two processes seem to occur. Sometimes the efficient acceleration even exhausts the rare $^3$He in the source region, limiting its fluence.

## 1    Introduction

The solar energetic particle (SEP) events observed first (Forbush 1946) were the largest and most energetic examples we know, where GeV protons produce a shower of particles cascading through the atmosphere to ground level in excess of that from the galactic cosmic rays (GCRs). However, these ground-level events (GLEs) provided little information on the composition of the incoming beam.

Measurement of multiple element abundances in SEPs began when Fichtel and Guss (1961) used nuclear emulsion detectors on a 4-min sounding-rocket flight from Ft. Churchill, Manitoba to observe a sampling of elements with $6 \leq Z \leq 16$. The principle elements up to the Fe abundance peak were observed in the next solar cycle by Bertsch et al. (1969) using the same technique. Resolution of He isotopes and the continuous time coverage needed to observe smaller events would only begin when detector telescopes were flown on satellites.

## 2    $^3$He

The first clearly-enhanced abundance ratio of $^3$He/$^4$He = $(2.1 \pm 0.4) \times 10^{-2}$ was reported by Hsieh and Simpson (1970), perhaps 50 times the value seen in the corona or solar wind, although differences in the spectra of the He isotopes were noted here.

Nearly every scientist involved in the early study of SEPs had previous experience with GCRs. After acceleration by shock waves at supernovae, GCRs spend ~$10^7$ years colliding with interstellar H to produce secondary $^2$H, $^3$He, and isotopes of Li, Be, and B. Thus the observation of $^3$He/$^4$He of 2% by Hsieh and Simpson (1970) was immediately misinterpreted as evidence that the SEPs had traversed enough material to fragment some $^4$He into $^3$He, just like the GCRs. However, it was soon found that there were many events like that seen by Serlemitsos and Balasubrahmanyan (1975) with $^3$He/$^4$He = 1.52 ± 0.10 but $^3$He/$^2$H > 300. How could there be more $^3$He than $^4$He and yet no $^2$H? Such abundances were definitely not compatible with fragmentation. Subsequently, there were limits established on Be/O and B/O in SEP events that were found to be < 2 × 10$^{-4}$ (e.g. McGuire et al., 1979; Cook et al., 1984). Thus these $^3$He-rich events were not just an accident of fragmentation; they must involve a completely new resonance phenomenon.

**Figure 1** shows a sample of $^3$He-rich events as we see them above 2 MeV amu$^{-1}$. As event intensities increase it becomes possible to see rarer ion species. Event 1 has only pre-event background levels of H, O, and Fe and $^4$He barely appears. $^3$He/$^4$He varies greatly in events and vs. energy, often peaking in the region 1 – 10 MeV amu$^{-1}$ (Mason, 2007), while Fe/O is much more stable once events are large enough to provide a measurable sample, so Fe/O is often used to define "impulsive" events (e.g. Reames et al., 2014). As discussed below, very high-Z elements begin to appear in the larger impulsive events and $^3$He fluence may be limited by depletion of $^3$He ions in the source volume.

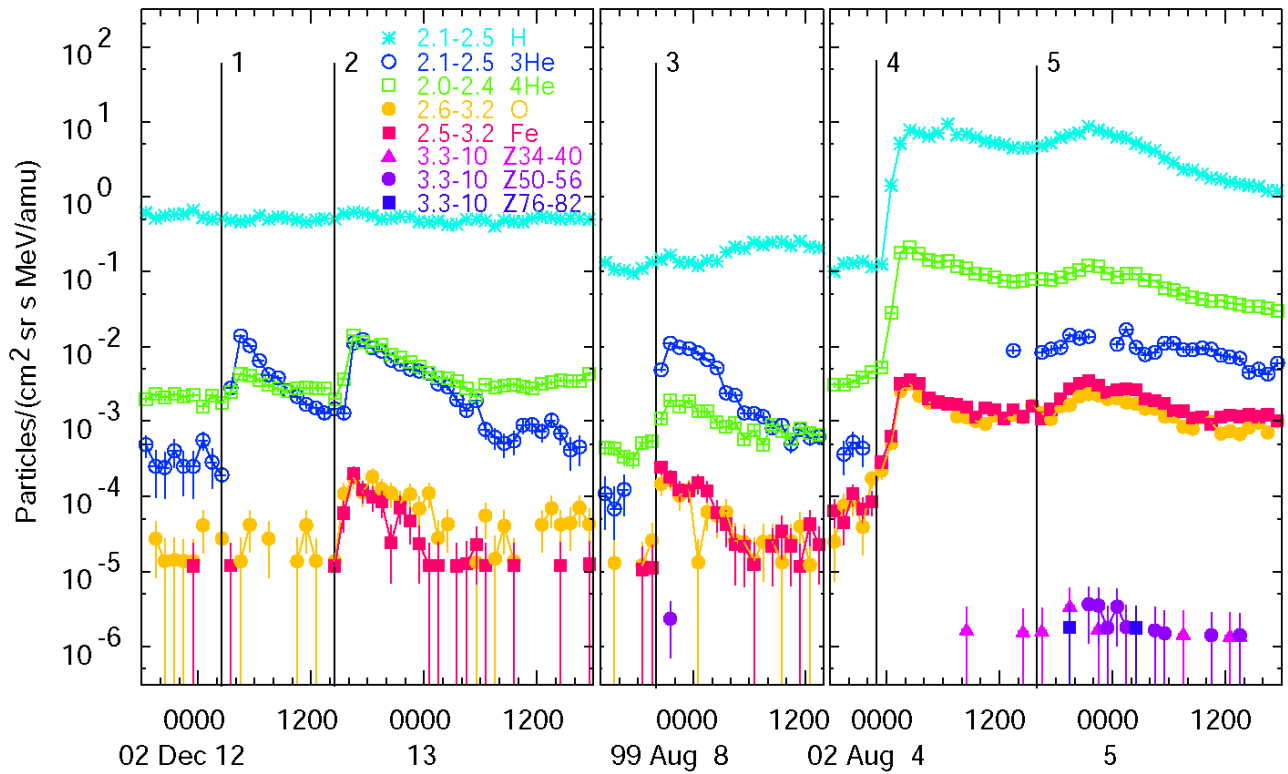

**Figure 1** Ion intensities at the indicated MeV amu$^{-1}$ are shown in a sample of $^3$He-rich or "impulsive" SEP events of increasing intensity. $^3$He exceeds $^4$He in Events 1 and 3. Fe and O are below background in Event 1, but Fe/O ≈ 1 in the remaining events. H is below background in Events 1 – 3. $^3$He/$^4$He < 0.1 in Event 4 and is omitted because of possible resolution errors; it is ≈ 0.1 in Event 5 and high-Z heavy elements are beginning to appear (Reames and Ng, 2004). Other properties of the last four of these events are listed in Reames et al. (2014a).

## 2.1 Electrons

Some of the earliest information on SEPs in space was provided by the radio emission produced by ~10 – 100 keV electrons streaming out from the Sun. The radio emission frequency varies as the square root of the local electron density which decreases sharply with distance from the Sun. In an early review article on solar radio bursts, Wild et al. (1963) distinguished type III radio bursts as produced by electrons streaming out rapidly along the magnetic field from a source at the Sun, and type II bursts which moved out at the slower ~1000 km s$^{-1}$ speed of an interplanetary shock wave.

Thus they saw two types of events: "electron events," which produced the type III bursts, and "proton events," accelerated at shock waves where associated electrons produce the type II burst as they are accelerated then overtaken, often at the flanks of the shock.

Early instruments flown on satellites measured 40 keV electrons associated with X-ray bursts at the Sun and type III radio bursts in space (Lin, 1970, 1974). These events were clearly different from the large proton events and some seemed to be "pure" electron events, i.e. lacking measurable ion intensities. The electrons events conformed to the picture presented by Wild et al. (1963).

It was not until 1985 that "pure" electron events turned out to be $^3$He-rich events (Reames et al., 1985). $^3$He-rich events were strongly associated with type III bursts, both the metric radio events, near the Sun (Reames et al., 1985), and the kilometric events below 2 MHz produced as the electrons continued out beyond ~6 solar radii (Reames and Stone, 1986).

## 2.2 Abundances of Elements

Measurements of element abundances soon began to show periods when Fe/O ≥ 1 (Mogro-Compero and Simpson, 1972; Gloeckler et al., 1975), an enhancement by a factor of ~10, relative to "coronal" abundances determined by the average of gradual SEP events (e.g. Reames, 1995a, 2014, 2021a), and such enhancements were shown to correspond with $^3$He-rich events (e.g. Mason et al., 1986), as seen in **Figure 1**. Reames (1988) looked at daily averages of SEPs to measure the overall distribution of abundances. He found a bimodal pattern with two branches of Fe/O. The branch near Fe/O ≈ 1 was $^3$He rich, electron rich, and proton (H/$^4$He) poor relative to the branch near Fe/O ≈ 0.1. These bimodal abundances were helpful in distinguishing the physics of impulsive and gradual SEP events.

Reames et al. (1994) found that on average in $^3$He-rich events, the elements He, C, N, and O were unenhanced relative to coronal abundances, Ne, Mg, and Si were enhanced a factor of ~2.5 and Fe was enhanced a factor of ~7. This pattern would occur if He – O were fully ionized and Ne, Mg, and Si were in a stable state with two orbital electrons, which occurs in the temperature range 3 – 5 MK. This suggested that patterns of element abundance enhancements could be used to determine source plasma temperatures, since the pattern of $Q$ values and $A/Q$, thus element enhancements, was dependent upon temperature (Reames et al., 2014b; Reames, 2018). The temperature variations among impulsive events turned out to be small, so the technique was more useful for gradual SEP events (Reames, 2016) with larger variations in temperature.

The extensive enhancement of very heavy elements was suggested early when Shirk and Price (1974) studied etch pits in a glass window of the *Apollo 16* lunar command module and found $(Z > 44)/Fe = 120^{+120}_{-60}$ at $0.6 \leq E \leq 2.0$ MeV amu$^{-1}$ from a small SEP event in April 1972. Routine measurement that resolved elements above Fe with $\delta Z/Z \sim 2\%$ began with the launch of the *Wind* spacecraft in November 1994. Reames (2000) found significant enhancements up to ($70 \leq Z \leq 82$), at $3.3 \leq E \leq 10$ MeV amu$^{-1}$, but only in impulsive SEP events. These measurements improved statistically with time (e.g. Reames and Ng, 2004) until Reames et al. (2014a) found the dependence of enhancements rising at the 3.64 ± 0.15 power of $A/Q$ (at ≈3 MK) from He to Pb, with the ($76 \leq Z \leq 82$)/O interval enhanced by a factor of ≈900. Below 1 MeV amu$^{-1}$, Mason et al. (2004) found enhancements varying as a power of $A/Q$ of 3.26 and the interval $180 \leq A \leq 200$ was enhanced a factor of ≈200. With the exception of $^3$He, enhancements are not strongly energy dependent.

When the power-law fits to abundance enhancements vs. $A/Q$ for elements $Z \geq 6$ in an event were extrapolated down to H at $A/Q = 1$ (Reames, 2019b), there were small impulsive SEP events that seemed to fit the protons extremely well (called SEP1 events; Reames 2020) and some larger events with a large proton excess (called SEP2 events) as shown by the examples in **Figure 2**.

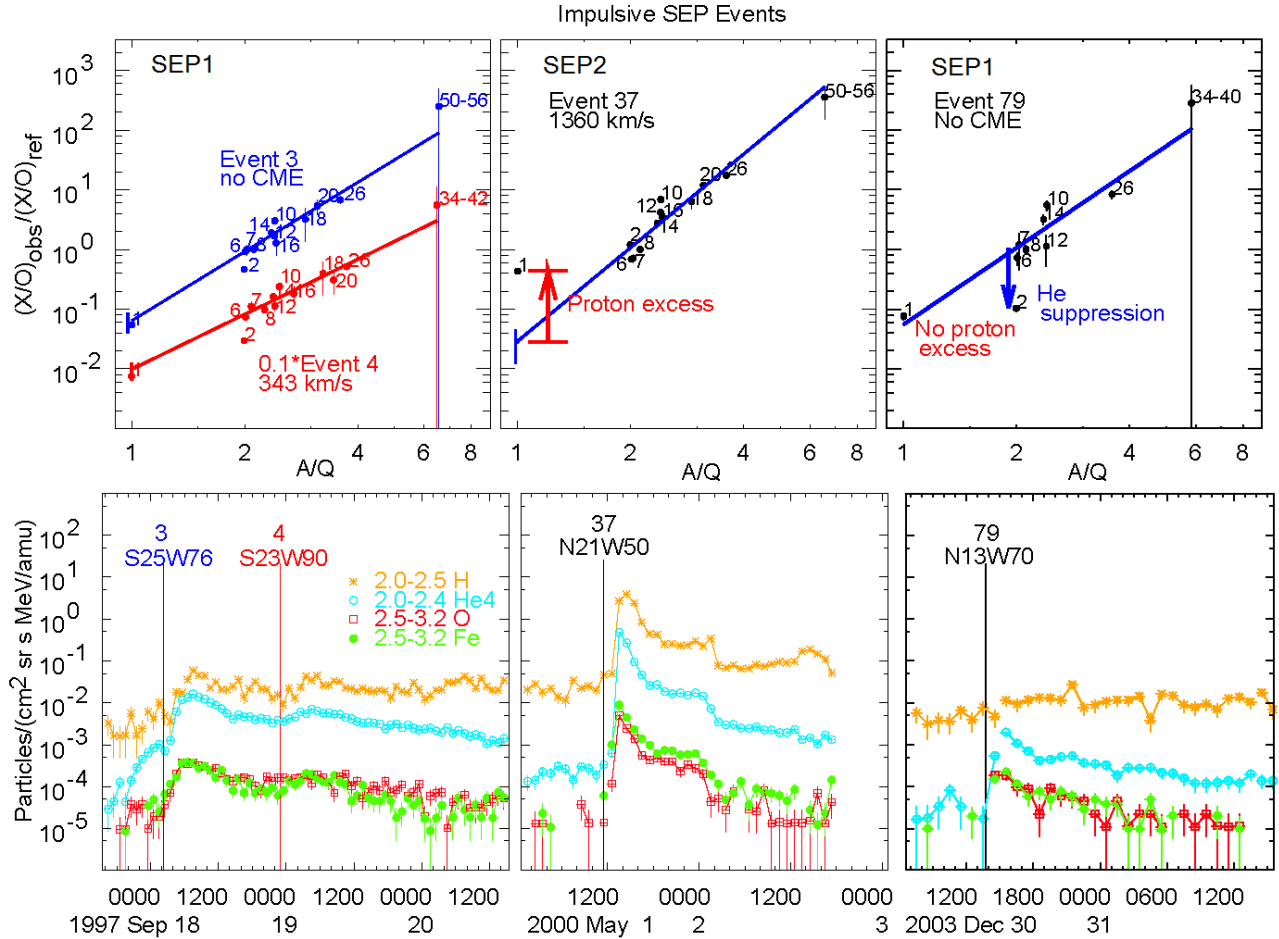

**Figure 2 Lower** panels show time histories of H, $^4$He, O and Fe, at the indicated MeV amu$^{-1}$, for four impulsive SEP events, two small SEP1 events on the **left** and a larger SEP2 event in the **center**, and a $^4$He-poor SEP1 event on the **right**. Event numbers shown above source coordinates refer to the event list of Reames et al. (2014a). Power-law fits to the abundance enhancements, noted by Z, in each event, are shown in the **upper** panels and CME speeds are listed when CMEs are seen. The **center** Event 37 is an SEP2 class event because of its large proton excess noted. For the SEP1 events on the **left** and **right**, the protons abundances lie on the extrapolated power-law fits (Reames, 2019b, 2020).

We will see below that the event in the **central** panels of **Figure 2** was one of the first impulsive events to be associated with a narrow CME produced by a solar jet (Kahler et al., 2001). These SEP2 events are generally more intense and were often associated with CMEs fast enough (1360 km sec$^{-1}$ in this case) to drive shock waves that could reaccelerate the SEP1 impulsive suprathermal ions from the earlier magnetic reconnection as well as ions, at least H ions and occasionally He, from the ambient plasma. The correlation of SEP proton intensity with proton excess is shown in **Figure 3** along with the suggested explanation of the excess.

At a temperature of 2.5 – 3 MK the elements $^4$He and C are both fully ionized with $A/Z = 2$, so $^4$He/C should represent the underlying coronal abundance, yet ~6% of impulsive SEP events are extremely $^4$He poor (Reames 2019a) with $^4$He/C ≈ 15 vs. an average 137±8. One such event is shown as Event 79 in the **right** panel of **Figure 2** where H, near background level, lies near the power-law fit of high-Z elements, but He lies well below it; another is shown as Event 3 in the **central** panel of **Figure 1**, where $^4$He is obviously much closer to O and Fe than it is in other events. It has been suggested that the high first ionization potential (FIP) of He, of 24.6 eV can delay its ionization and elevation into the corona, but Ne with FIP = 21.6 eV is unaffected. Any possible $Z^2/A$ effects from matter traversal would seem to suppress C more than $^4$He. What causes this occasional $^4$He poverty?

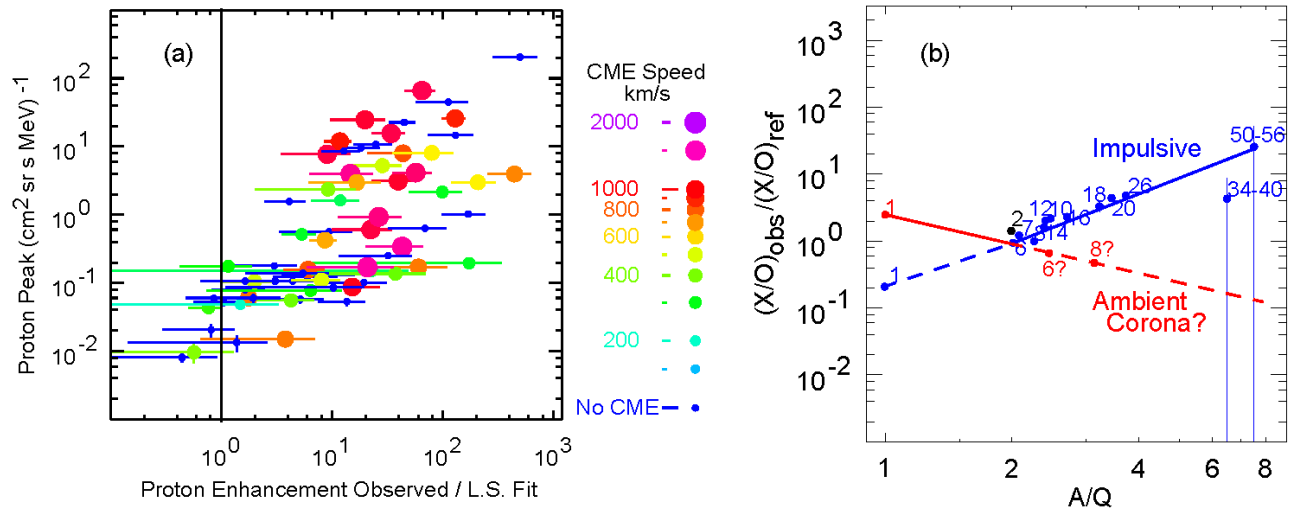

**Figure 3** Panel **(a)** shows the peak proton intensity at 2 – 2.5 MeV vs. the proton excess relative to the $Z >2$ power-law fit for impulsive SEP events with the symbol size and color determined by CME speed as shown. Events with fast CMEs have proton excesses. Panel **(b)** suggests two possible contributions to the plot of enhancement vs. $A/Q$ for the shock-enhanced SEP2 impulsive events: reaccelerated impulsive SEP1 seed particles (**blue**) and accelerated ambient coronal seed particles (**red**). Note that a shock wave can accelerate outside the volume that contains the SEP1 seed particles. If the shock is very strong, the ambient corona could dominate and the event could become a shock-dominated gradual event.

Note that we have studied variations of H/O and $^4$He/O separately in the above and *not* H/$^4$He. Thus in Figure 2 we identify separate processes where H variation from the extrapolated power-law fit of $Z > 2$ depends upon the presence and activity of fast shock waves, while $^4$He (presumably) depends upon the especially high FIP of He. Studying the ratio H/$^4$He would have blurred together the effects of these extremely different physical processes. This important feature was not initially obvious.

Isotope resolution has been extended as high as Fe (e.g. Leske et al. 1999, 2007). These measurements show variations in $A/Q$ that supplement similar measurements that have been shown with power-law fits using multiple elements.

## 2.3 Ionization States

Luhn et al. (1984, 1987) provided important early measurements of ionic charge states, $Q$, up to Fe for energies 0.34 – 1.8 MeV amu$^{-1}$. For large gradual events they found average values of $Q_{Si}$ = 11.0± 0.3 and $Q_{Fe}$ = 14.1 ± 0.2, appropriate for a temperatures of ≈ 2 MK. However, for $^3$He-rich events they found $Q_{Si}$ ≈ 14 and $Q_{Fe}$ = 20.5 ± 1.2, which meant either a source temperature of ≈10 MK, or stripping of the ions after acceleration. How could the average abundances of Ne, Mg, and Si get enhanced if they were all fully ionized with $A/Q$ = 2 during acceleration, just like He, C, and O? This dilemma was finally resolved when DiFabio et al. (2008) found that the ionization states in the impulsive SEP events increased with ion speed, suggesting that the ions had come to ionization equilibrium during traversal of a small amount of material *after* acceleration. DiFabio et al. (2008) concluded that acceleration must have occurred near 1.5 solar radii.

## 2.4 Acceleration Theory

Most of the early attempts to explain the enhancement of $^3$He involved the preferential absorption of some form of wave energy, in resonance with the gyrofrequency of $^3$He, to produce selective preheating that would enhance the tail of thermal distribution of $^3$He so as to inject more ions into some unspecified acceleration mechanism. Ibragimov and Kocharov (1977) and Kocharov and Kocharov (1978, 1984) were first, considering ion-sound wave heating, but Weatherall (1984) pointed out that this could not account for all abundances. Fisk (1978) and Varvoglis and Papadopoulis (1983) suggested selective heating by absorption of electrostatic ion cyclotron waves. Winglee (1989) considered the ion-ion streaming instability to enhance heavy ions, and Riyopoulos (1991) considered electrostatic two-ion (H–$^4$He) hybrid waves.

Temerin and Roth (1992) considered the ubiquitous associated streaming electrons that produced electromagnetic ion cyclotron (EMIC) waves that were adequate to actually resonantly accelerate the $^3$He that absorbed the waves while mirroring in the magnetic field, in analogy with "ion conics" seen in the Earth's aurorae. Roth and Temerin (1997) suggested that heavier ions were enhanced through their second harmonic, but there was no smooth power law in $A/Q$. Miller et al. (1993a, b) considered the effects of other electron-beam-generated wave modes and Steinacker et al. (1997) considered warm-plasma broadening of spectral lines to produce a "He valley" of wave absorption.

Litvinenko (1996) considered effects of EMIC waves and Coulomb energy losses on the $^3$He spectrum and Liu et al. (2004, 2006) were able to fit the complex spectra of $^3$He and $^4$He with a model of stochastic acceleration by a power-law spectrum of plasma-wave turbulence. Roughly speaking, as the ions are accelerated to higher energy, the $^3$He begins to form a distinct energy peak as the $^3$He in the source volume becomes depleted, while the $^4$He does not.

Separately, explanation of the power-law dependence of enhancements on $A/Q$ was first found by Drake et al. (2009) in particle-in-cell simulations of collapsing islands of magnetic reconnection. Ions are Fermi accelerated as they scatter back and forth from the ends of the collapsing islands. A similar process is found to accelerate electrons (Arnold et al., 2021) and its efficiency depends upon the strength of the out-of-plane guide field. However, $^3$He has not been discussed at all in this process, although there would seem to be abundant opportunities for mirroring ions to absorb waves.

There is evidence that the contribution of $^3$He is often saturated, i.e. all the $^3$He in the accelerating volume is actually accelerated. Reames (1999) estimated the total number of energetic $^3$He ions integrated over energy, space, and time and found it could be comparable with the number that exist in a typical flare volume. Subsequently, Ho et al. (2005, 2019) found that this effect limited the maximum possible fluence of $^3$He. It has been suggested (Kahler, private communication) that comparing the calculated number of SEP $^3$He with the calculated number of $^3$He in the observed volume of a jet or reconnection region, on an event-by-event basis, might help determine where and what fraction of the $^3$He is accelerated. Is there enough $^3$He in the reconnection volume or is it accelerated throughout the jet as the Temerin and Roth (1992) model may suggest? Can anything be said about possible second-harmonic acceleration of some heavy ions like Fe, or the rare, low-energy resonant peaks of Si or S (Mason et al., 2016), for which the low-energy spectra roll over like that of $^3$He? Is SEP output bounded by the reconnection volume? Accelerating volumes are limited in impulsive SEPs while, in contrast, in gradual SEP events, source shock waves of huge area sweep out extensive volumes.

## 3    Flares, CMEs, Shocks, and Jets

Reconnection of the solar magnetic field drives nearly all solar activity, and flares provide a ubiquitous visible marker of magnetically trapped heating from that reconnection. Perhaps it is not surprising that the earliest evidence of GLEs was associated with obvious, bright flares, but this association was taken much too literally.

Radio data were the first to distinguish two acceleration mechanisms for SEPs in space (Wild et al., 1964), type III bursts that associate with electrons streaming from open magnetic reconnection in impulsive SEP events, and type II bursts that associate with shock acceleration and large gradual SEP events, but, unfortunately, the radio evidence was largely ignored in early SEP history.

After CMEs were identified and their observation became common, Kahler et al. (1984) found a 96% correlation between large SEP events and fast, wide CMEs. Shock waves driven by the CMEs could explain the extremely broad longitude span of the events. Mason et al. (1984) found that the minimal rigidity dependences of abundance variations across longitude were inconsistent with a point source origin and they discussed large-scale shock acceleration that they labelled LSSA. The importance of CMEs and shocks was not taken seriously by some flare enthusiasts until Gosling's (1993, 1994) review article entitled "The solar flare myth" was published. This article was then found to "wage an assault on the last 30 years of solar-flare research" (Zirin, 1994) even though most published "solar-flare research" ignored SEPs entirely. However, the importance of CMEs and the existence of two mechanisms of SEP acceleration began to be recognized (Reames, 1988, 1995b, 1999, 2013, 2015, 2020, 2021a, 2021b; Zank et al., 2000, 2007; Kahler, 2001; Cliver et al., 2004; Lee, 2005; Lee et al., 2012; Gopalswamy et al., 2012; Mewaldt et al. 2012; Desai and Giacalone, 2016; Kouloumvakos et al., 2019).

For a brief interval, distinguishing impulsive and gradual SEP events seemed simple. Impulsive events were $^3$He-rich and gradual events were not. Then Mason et al. (1999) found a small but significant enhancement of $^3$He in a large SEP event that would be called gradual in all other respects. It became clear that shock waves in gradual SEP events could reaccelerate residual suprathermal ions left over from previous impulsive SEP events. In fact these pre-accelerated ions might be preferred in some cases, as in quasi-perpendicular shocks when ions needed to overtake the shock from downstream (Tylka et al., 2001, 2005; Tylka and Lee, 2006). In fact, it became evident that large pools of $^3$He-rich, Fe-rich suprathermal ions were extremely common, and available for shocks to traverse (Richardson et al., 1990; Desai et al., 2003; Wiedenbeck et al., 2008; Bučík et al., 2014, 2015; Chen et al., 2015). Whenever there was no large SEP event in progress the default suprathermal ion abundances below ~1 MeV amu$^{-1}$ seemed to be $^3$He-rich and Fe-rich, suggesting a large number of small unresolved jets (nanojets?) could generate SEPs faster than the solar wind could sweep them away. Thus, for 24% of gradual events, the $Z \geq 2$ elements are dominated by reaccelerated ions from impulsive events, called SEP3 events (Reames, 2020), and in 69% of gradual events the shock predominantly accelerates ions, even $Z > 2$ ions, from the ambient coronal plasma (SEP4 events; Reames, 2020, 2021a, 2021b).

Impulsive SEP events were small and difficult to associate with coronal features, but Kahler et al. (2001) were able to associate several of the larger impulsive SEP events with narrow CMEs. **Figure 4** shows a narrow (54º), fast (1360 km s$^{-1}$) CME from the impulsive event of 1 May 2000 SEP, seen by the Solar and Heliospheric Observatory (SOHO; https://sohowww.nascom.nasa.gov/), for which SEPs were shown in the **central** panel of **Figure 2**. Such CMEs had been associated with type III bursts and solar jets (Shimojo and Shibata, 2000). Unlike flares, jets involve magnetic reconnection on open field lines so the SEPs (and the CMEs) easily escape. Tracking of impulsive SEPs back to the Sun led to jets, often on the boundary between active regions and coronal holes (Wang et al., 2006; Nitta et al. 2006, 2015). However, onset times are poorly defined in small SEP events, making associations with type III bursts, etc. more difficult. Reconnection is often triggered by large-scale waves moving across the corona (Bučík et al., 2016). It is now possible to consider the nature of the associated jets directly for many $^3$He-rich events (Bučík et al., 2018a, 2018b, 2021; see review Bučík, 2020). It has even been possible to compare the temperatures derived from the extreme ultraviolet (EUV) images of coronal source regions of 24 solar jets associated with ³He-rich events (Bučík et al., 2021) with the abundance-derived temperatures from SEPs (Reames et al., 2014b).

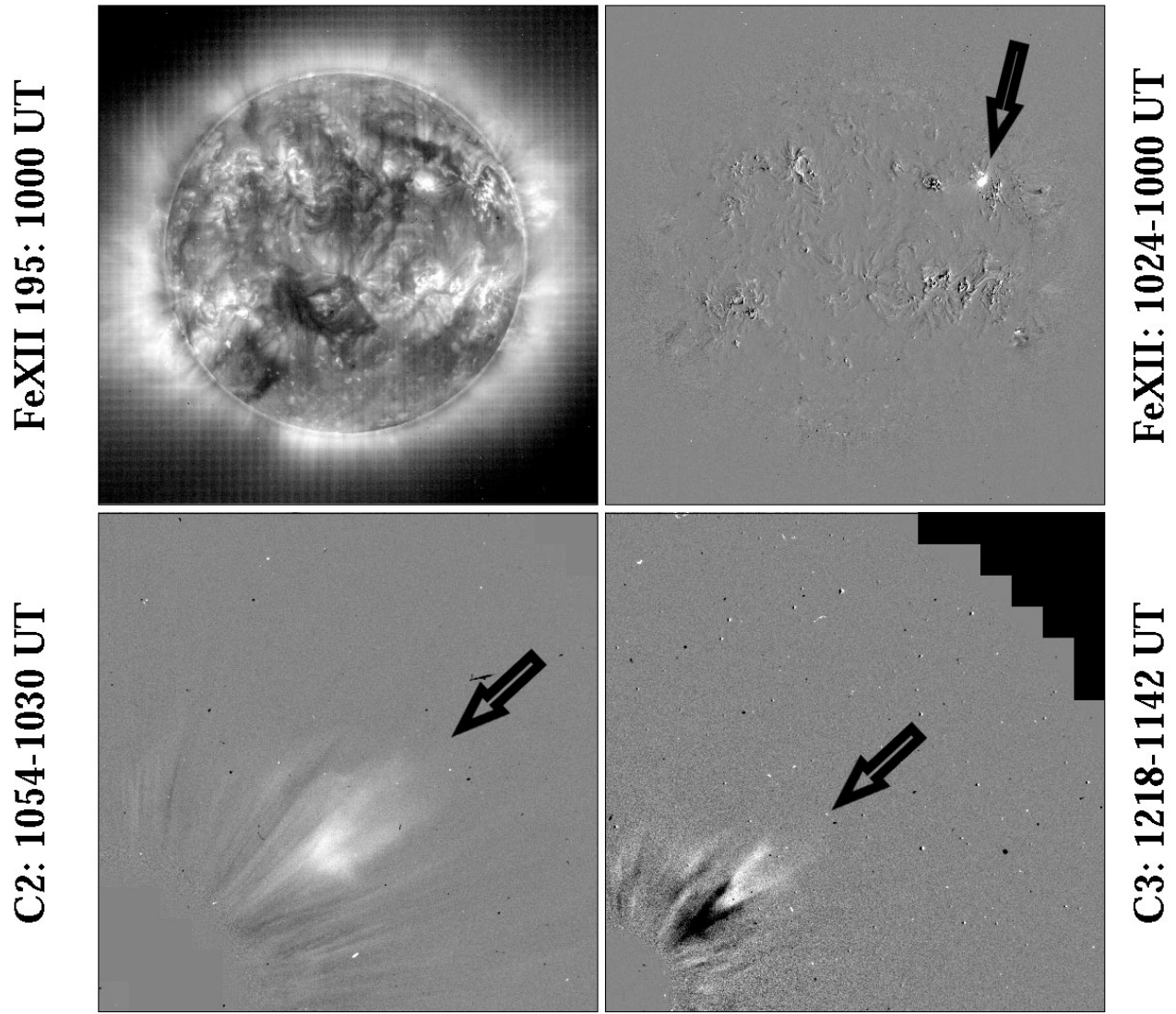

**Figure 4** Images of the impulsive SEP event of 1 May 2000 (see also **Figure 2**) are: **upper left** a full disk SOHO/EIT image at 195 Å *before* the flare, **upper right** subtracted EIT image at 195 Å with arrow showing new small white source in the NW quadrant, **lower left** subtracted image of NW quadrant of SOHO/LASCO/C2 coronagraph 2 – 6 $R_S$ with arrow showing narrow CME, **lower right** subtracted NW quadrant of C3 image 4 – 30 $R_S$ with arrow showing narrow CME (Kahler et al. 2001). Tracking and magnetic configuration for this event are shown by Nitta et al. (2006) and by Wang et al. (2006).

However, impulsive SEP events are not necessarily derived purely from magnetic reconnection; the CMEs from these jets are often fast enough to drive shock waves which can reaccelerate SEPs from the reconnection as is certainly the case for the 1360 km s$^{-1}$ CME in the 1 May 2000 event shown in **Figure 4**. It may be possible to distinguish the pure (SEP1) events from those reaccelerated by a shock (SEP2) from the proton excess produced in the latter events as shown in **Figure 3**.

A sketch showing a jet formed by newly emerging magnetic field is shown in **Figure 5**. When the emerging field has opposite polarity from that of the preexisting field, reconnection takes place, not at a single point but in a series of "islands." In realistic jets, the opposing fields rarely exactly cancel, but rather leave an out-of-plane, residual "guide field." Particle acceleration occurs as particles, mainly electrons (Arnold et al. 2021), are Fermi-accelerated as they pitch-angle scatter in the evolving fields. As the SEPs and CME plasma are ejected on open field lines at the upper right in **Figure 5**, newly closing loops capture some SEPs in the lower left region labeled "flare" where they deposit their energy as heat. While much of the jet, including the SEPs, retain temperatures of 2 – 3 MK and emit EUV (Reames et al., 2014b; Bučík et al., 2021), the "flare" region with its captured SEP energy heats to 10 – 20 MK and emits X-rays. These X-rays often provide the source location. It is important to realize that the presence of the X-rays does not define the nature of the SEPs in space; both environments coexist since the reconnection that opens some field lines closes others. Some day someone will be able to provide us with an overlay of X-ray and EUV images that will map details of these source regions. As reconnection events become larger and more complex, both X-ray and EUV regions increase, contributing to a "big flare syndrome" (Kahler 1982), i.e. a misleading correlation between X-rays and the SEPs in space.

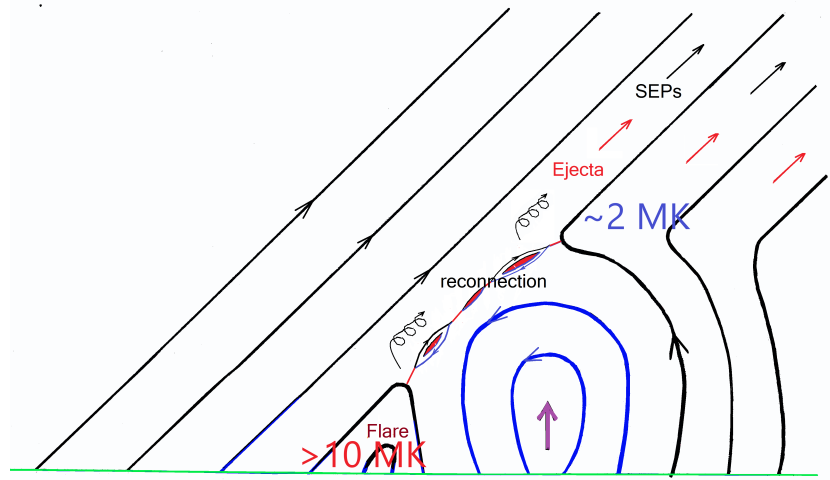

**Figure 5.** A jet is produced when newly emerging magnetic flux (*blue*) reconnects with oppositely directed field (*black*) in the *red* region. This reconnection region is not a uniform surface but forms multiple islands of reconnection. Energetic particles and plasma can escape toward the *upper right* and a newly-enclosed flaring region labeled "Flare" forms at the *lower left* that is heated by trapping SEPs. Real jets can be much more complex, involving twisted fields, etc.

What about solar flares? There is evidence from measurements of Doppler-broadened $\gamma$-ray lines that the ions accelerated in large solar flares are $^3$He-rich (Mandzhavidze et al., 1999; Murphy et al., 2016) and Fe-rich (Murphy et al., 1991), just like the impulsive SEPs from jets that we see in space. The SEPs accelerated on closed loops, dominated by electrons, soon scatter into the loss cone and plunge into the denser corona below, scattering against ions to produce X-ray bremstrahlung, and heating the plasma which expands back up into the loops creating a hot (>10 MK), bright flare. SEPs from jets are not hot (~2 – 3 MK; Reames et al., 2014b; Bučík et al., 2021) because the SEPs and the CME plasma escape. Flares exist precisely because their energy is magnetically trapped and can only escape as heat, light, or neutral particles. Reconnection of closed fields with other closed fields cannot produce open fields, except when those fields are eventually carried outward by a CME.

Shock waves, driven by fast, wide CMEs are the basis of gradual SEP events. However, the narrow (<60º) CMEs emitted from solar jets are also fast enough to drive significant shocks, such as the 1360 km s$^{-1}$, 54º width CME in the 1 May 2000 events discussed above. Yet, often, these fast, narrow CMEs from jets lack type II radio bursts that can signify shock acceleration, and the CME's contribution to SEP acceleration may be questioned (e.g. Bronarska 2018; Kahler et al. 2019). At quasi-parallel shock waves, ions are accelerated by scattering back and forth across the shock against Alfvén waves (often self-generated), but non-relativistic electrons cannot resonate with Alfvén waves, so they can only be accelerated in the $\mathbf{V_S} \times \mathbf{B}$ electric field in more quasi-perpendicular regions of the shock. A big wide hemispherical shock surely has some regions where type II electrons get accelerated, while a narrow CME may produce only quasi-parallel regions – accelerating few electrons but plenty of ions. The shock from the 1 May 2000 event may not accelerate enough electrons, but it surely accelerates more ions than any of the shocks we have been able to observe directly in situ with shock speeds as low as 300 km s$^{-1}$ (Reames 2012). The one feature that seems to show the presence or absence of shock acceleration in impulsive SEP events is the proton excess shown in Figures 2 and 3 above (Reames 2019b), but the response of electrons and protons can be completely different.

Some people continue to defined "flare" to include every energetic phenomenon on the Sun, including flares, jets, CMEs, shocks, and even the SEPs at 1 AU. That, of course, makes flare research the most important discipline of all, studying the cause of everything solar, by definition. This deliberately blurs the physics, elevating the importance of flares and diminishing that of CMEs; this was exactly Gosling's (1993) objection to "The Solar Flare Myth." Applying "flare" to everything is not only meaningless but helps no one understand any physics. Beginning with Carrington (1960), many of us still think of "the flare" as that localized sudden bright flash of white light, Hα, or X-ray emission driven because reconnection energy is trapped in closed magnetic loops on the Sun – an event recorded by the Solar Flare Patrol and documented by its timing, latitude, longitude, and C-M-X-scale soft X-ray intensity. These flares are limited in spatial extent. In contrast, the shock wave that accelerates SEPs in a large gradual event is a nominally hemispherical structure initially active from 2 to over 3 R$_S$ (Reames 2009a, b; Cliver et al. 2004), thus enclosing an accelerating volume perhaps ~10 times the volume of the Sun itself – this is *not* a flare. Thus the source of SEPs has a major effect on their spatial distributions, among many other things, and lumping all possible sources together, to inflate the egos of a few flare researchers, is no help. Please distinguish CMEs from flares in your publications.

This field began with a frequent assumption that all SEPs were somehow actually accelerated in flares. First we found that SEPs in gradual events were accelerated by CME-driven shocks instead. Then we found that the remaining impulsive SEPs in space came from jets. Now we find that it may be more correct to say that flares are caused by SEPs trapped on loops than the converse.

## 4 Conflict of Interest

*The author declares that this research was conducted in the absence of any commercial or financial relationships that could be construed as a potential conflict of interest*.

## 5 Author Contributions

All work on this article was performed by D.V. Reames.

## 6 Funding

No institutional funding was provided for this work.

## 7 Acknowledgments

The author thanks Steve Kahler for helpful comments on this article.